\documentclass[jcp,twocolumn,showpacs,showkeys,preprintnumbers,amsmath,amssymb]{revtex4}
\usepackage{graphicx}
\usepackage{dcolumn}
\usepackage{bm}
\newcommand{\ket}[1]{|#1\rangle}  
\newcommand{\bra}[1]{\langle#1|}  
\newcommand{\brkt}[2]{\langle#1|#2\rangle}  
\newcommand{\eq}[1]{Eq.~(\ref{#1})}  
  
\newcommand{\be}[1]{\begin{equation}\label{#1}}  
\newcommand{\ee}{\end{equation}}  
  
\begin{document}

\title{Stochastic dissociation of diatomic molecules}   
\author{Anatole Kenfack}  
\email[Corresponding author:]{kenfack@pks.mpg.de}   
\author{Jan M. ~Rost} 
\affiliation{Max Planck Institute for the Physics of Complex Systems, N\"othnitzer Strasse 38, 01187 Dresden, Germany. }


\begin{abstract}  
The fragmentation of diatomic molecules under a stochastic force is  
investigated both classically and quantum mechanically, focussing on  
their dissociation probabilities.  It is found that the quantum system is more robust than the  
classical one in the limit of a large number of kicks.  
The opposite behavior emerges for a small  number of kicks.  Quantum and classical dissociation probabilities 
do not coincide for any parameter combinations of the force.   This can be attributed to a scaling property  in the classical system which is broken  quantum mechanically.    
\end{abstract} 

\pacs{33. 80. Gj, 34. 10. x, 03. 65. Sq} 
\keywords{diatomic molecules, stochastic force, Wigner  
  function, dissociation probablity, noise. }  

\maketitle  
 
\section{Introduction}  
The anharmonicity of molecular vibrations makes the dissociation of a molecule by irradiation of laser light a relatively difficult task  \cite{bloembergen}.  Consequently, high intensity is required for dissociation, for instance, $I>10^{15}W/\mathrm{cm}^2$  
for $HF$ and $I\ge10^{14}W/\mathrm{cm}^2$ for $HCl$.  At such intensities, however, the ionization process dominates and masks vibrational excitation and dissociation.  Chelkowsky et al. ~ \cite{cheal90} suggested that the dissociation threshold of a diatomic molecule can be lowered by two orders of  
magnitude using a frequency chirped laser, and hence dissociation without ionization should be possible.  In a similar   
spirit, circularly chirped pulses have been used by Kim et al. ~ \cite{kim} for the dissociation of  
diatomic molecules.  They found that the threshold laser intensity is sufficiently  
reduced, to achieve  dissociation without ionization.    
  
Here, we investigate the possibility for dissociation of diatomic molecules under a stochastic force, which could eventually be chosen such that ionization is minimal.  A second motivation for our work is the question, if a stochastic driving destroys quantum coherence and eventually brings the quantum and classical evolution close to each other.   
  
We model the force as a sequence of pulses  
(kicks) at random times, each kick carrying an independent  
weight~ \cite{masoliver}.  This type of force, similar to white shot noise,   
has been used to model the passage of ions through carbon foils before \cite{burgdorfer}.   Its average strength $\gamma$ and the average  
number of kicks $\langle \Delta t \rangle$ determine the  
dynamics of the system which is taken as a Morse oscillator 
\cite{morse} with parameters corresponding to Hydrogen Flouride (HF) 
and Hydrogen Chloride (HCl) molecules.   
 Classical and quantum evolution of the system are conveniently compared by using the Wigner transform  
of the initial wavefunction as initial phase space distribution in the classical evolution.   
  
 We begin the main text of this paper in section II with a brief 
description of the stochastic Hamiltonian.  
In section III we explain the classical and quantum method with which we solve the stochastic dynamics, which is the Langevin equation with test 
particle discretization and Monte Carlo sampling in the classical case and the 
direct solution of the  
stochastic Schr\"odinger equation with a standard FFT Split Operator method with absorbing boundary conditions 
in the quantum case. Results, particularly for the dissociation probability will be presented and discussed in section IV, while section V concludes the paper.  
  
\section{Description of the Model}  
  
The one-dimensional stochastic Hamiltonian of our system reads  
(atomic units are used unless stated otherwise)  
\begin{equation}  
H(t)=H_{0}-xF(t)\equiv \frac{p^{2}}{2m}+V(x) -xF(t),  
\label{hamiltonian}  
\end{equation}  
where the molecular dipole gradient \cite{cheal90} has been absorbed into the stochastic force $F(t)$.    The Hamiltonian $H_{0}$ describes vibrational motion of the molecule  in the Morse potential  \cite{morse}  
\begin{equation}  
V(x)=-D_{e}+D_{e}\left( 1-\exp(-\beta x)\right)^{2}  
\label{potential}  
\end{equation}  
with well depth $D_e$ and and length scale $\beta$.   
The eigenenergies $E_n$ of the Morse oscillator $H_0$ are given by  
\be{eigenvalues}  
E_n=\hbar\omega_e(n+1/2)[1-B(n+1/2)/2],\,\,\,\,\,\,\,\,\,0\le n\le [j]  
\end{equation}  
where $\omega_e$ is the harmonic frequency, $n_b = [j]+1$ is the number of bound states with  
\be{parameters}  
j=1/B-1/2,\,\,\,B = \hbar\beta(2mD_e)^{-1/2},\,\,\,\hbar\omega_e=2BD_e\,.  
\end{equation}  
The parameters specific to HCl and HF are given in Table \ref{table1}. 
\begin{table}
\begin{tabular}
{c|c|c|c|c|c}
Molecule&B &$D_e$ [eV] &$\beta$ [$a_0^{-1}$]&
$N_b$&$\omega_e$ [Hertz]\\\hline\hline
HCl&$4. 07\times 10^{-2}$&4. 40&0. 9780&25&$8. 66\times 10^{13}$\\\hline
HF&$4. 19\times 10^{-2}$&6. 125&1. 1741&24&$12. 38\times 10^{13}$\\\hline
\end{tabular}
\caption{Parameters of  the HF and HCl molecule for the Morse potential. }
\label{table1}
\end{table} 
The stochastic force $F(t)$  \cite{masoliver,haenggi} in   
\eq{hamiltonian}  
\begin{eqnarray}  
F(t)=\sum_{i=1}^{N_t}\gamma_{i}\delta(t-t_{i}),\label{force}  
\end{eqnarray}  
stands for a series of random impulses of strength $\gamma_{i}$ at times $t_{i}$, i. e  
$F(t)$ is a kind of white shot noise  \cite{broeck} responsible for multiple $\delta-$kicks undergone by the molecule,  
 where $N_t$ is the number of kicks up to time $t$ controlled by the Poisson counting process $N_t$.  It is characterized by the average kicking interval $\langle \Delta t\rangle \equiv \lambda^{-1}$ about which the actual interval $\Delta t_i=t_i-t_{i-1}$ are exponentially distributed, similarly as the actual kicking strenghts $\gamma_i$ about  
their mean $\gamma$,  
\be{deltat}  
P(\Delta t_i) = \lambda\exp(-\lambda\Delta t_i),\,\,P(\gamma_i) = \gamma^{-1}\exp(-\gamma_i/\gamma)\,.   
\ee  
  
 We restrict our analysis to positive $\gamma_{i}$  
and assume that $\gamma_{i}$ and $t_{i}$ are mutually uncorrelated random variables generated by the distributions functions of \eq{deltat}.  The determination of $F(t)$ reduces to the construction of a stochastic sequence $(\gamma_{i},t_{i})$ which can be done assuming that the random times $t_{i}$ form a Poisson sequence of points leading to a delta correlated process \cite{masoliver}.    It is easy to show \cite{haenggi}  that the constructed stochastic force  
has the following properties: \begin{eqnarray} \langle F(t) \rangle&=&\gamma\lambda \nonumber\\ \langle F(t)F(s)\rangle&=&2\gamma^2\lambda\delta(t-s)+\gamma^2\lambda^2\,.  \label{averageforce} \end{eqnarray} The corresponding power spectrum, i.~e. , the Fourier transform of $\langle F(t)F(s)\rangle$, is given by  \begin{eqnarray} S(\omega) = 4\frac{\gamma^2\lambda}{\sqrt{2\pi}}+\gamma^2\lambda^2\sqrt{2\pi}\delta(\omega).  
\label{powerspectrum} \end{eqnarray} These properties reveal  the difference between the present stochastic force (white shot noise) and a pure white noise which is delta-correlated with {\it zero mean}.   
 
\section{Dynamics}  
\subsection{Time evolution}  
Our system as described is non deterministic due to the stochastic nature of its Hamiltonian, but closed.  This is consistent with a regime of high effective temperature and no dissipation.  Specifically speaking, the system is a simple anharmonic particle which is not coupled to any environment (zero dissipation and no diffusion) but subject to an external force \cite{leggett}.  A  perturbative solution for this system is in general not possible, because the field strengths applied  significantly distort the system.    
We are interested in formulation of the dynamics which  
is applicable for the  quantum as well as for the classical treatment.   
This can be done in an elegant way by propagating the Wigner transform $W$ of the density matrix $\rho (t)= \ket{\psi(t)}\bra{\psi(t)}$ with a (quantum) master equation \cite{zurek} \begin{eqnarray} i\hbar\frac{\partial W}{\partial t}&=&L_{cl}W+ L_qW +\hat{O}W\nonumber\\ L_{cl}&=&-\frac{p}{m}\frac{\partial }{\partial x}+\frac{\partial V}{\partial x}\frac{\partial }{\partial p}, \nonumber\\ L_q&=&\sum_{n (odd)\ge 3}^{\infty}\frac{1}{n!} \left(   \frac{\hbar}{2i} \right)^{n -1}\frac{\partial^n   V}{\partial x^n}\frac{\partial^n} {\partial p^n}\,. \nonumber \end{eqnarray} Here, $L_{cl}$ and $L_q$ represent the classical and quantum Liouville operators, respectively, while $\hat{O}$ stands for the superoperator resulting from random kicks undergone by the molecule.    
Unfortunately, solving the master equation and constructing  $\hat{O}$ is a complicated task.   It is much   
easier to solve the equations of motion derived from  \eq{hamiltonian}
directly. 

The classical time evolution obeys the Langevin equation 
\begin{eqnarray} \frac{dp}{dt}=-\frac{\partial V}{\partial
    x}+F(t),\label{cl_langevin} 
\end{eqnarray}
while the quantum evolution can be obtained from the stochastic Schr\"odinger equation 
\begin{eqnarray} i\hbar\frac{\partial
    \ket{\psi(t)}}{\partial t}=H(t)\ket{\psi(t)}\,\label{q_langevin}.
\end{eqnarray} 
Both formulations have in common that they must be solved over a
larger number of realizations $N_r$ of the stochastic force.  Only the
average over all realizations produces the solution of the Classical
Langevin and the stochastic Schr\"odinger equations, respectively.   
  
\subsection{Initial state}  
  
 The molecule is  considered to be initially in the ground vibrational state $|\psi_0\rangle \equiv |0\rangle$ with  energy $E_0$ (\eq{eigenvalues}).  For the classical propagation we take the  Wigner distribution of the ground state as initial phase space distribution.   
Analytically, the initial phase space density is given by  
\begin{eqnarray} W_0(x(\xi),p)=\frac{2}{\pi\hbar\Gamma(2j)}\xi^{2j}K_{-\frac{2ip}{\hbar\beta}}(\xi)\,,   
\label{w_groundstate} \end{eqnarray}  
where $\xi(x)=2/B\exp(-\beta x)$ and $K_{\alpha}$ is the modified 
Bessel function of the third kind \cite{alejandro}. 

\subsection{Classical approach} The stochastic Langevin equation ($\ref{cl_langevin}$) can be solved numerically  
with test particles ("test particles discretization") so that
the Wigner function is given by
\be{monte}
W_(x,p,t)=N^{-1}_\mathrm{test}\sum_{k=1}^{N_\mathrm{test}}\delta (x-x_k(t))\delta (p-p_k(t))\,,
\ee
where $N_\mathrm{test}$ is the number of test particles and the
$(x_{k}(t),p_{k}(t))$ are the classically evolved trajectories of the test particles.
Their initial conditions $(x_{k}(0),p_{k}(0))$ are Monte Carlo sampled by dividing the phase space into small bins  $S_{lm}$~\cite{footnote}. In each $S_{lm}$ the initial conditions for $n_{lm}$
test particles are randomly chosen where $n_{lm}$ is determined by the value of the Wigner function $W(x,p)$ attached to the respective phase space bin,
\be{bin}
n_{lm} = W(x_l,p_m)S_{lm}N_\mathrm{test}
\ee
with $S_{lm} = (x_{l+1}-x_l) \times (p_{m+1}-p_m)$.

For each realization $r$ of the stochastic force \eq{monte} yields the 
propagated Wigner function $W_r(x,p,t)$ which must be averaged over the $N_r$ realizations to obtain the final result 
$W(x,p,t)$.
  
\subsection{Quantum approach}  
 For a given realization $r$, the solution of the stochastic Schr\"odinger equation (\ref{q_langevin}) amounts to solve the standard time-dependent  
Schr\"odinger equation \begin{eqnarray} \ket{\psi_{r}(t)}=U_{r}(t,t_{0})\ket{\psi(t_{0})}, \label{e36} \end{eqnarray} where $U_{r}(t,t_{0})$ is the evolution operator  and $|\psi(t_0)\rangle = |0\rangle$.  Since the stochastic force  
consists of instantaneous kicks,  $U_{r}(t,t_{0})$ can be written as \be{e38} U_{r}(t,t_{0})=U_0(t,t_{N_r})\prod_{i=0}^{n_r-1}\exp\left( \frac{i}{\hbar}x\gamma_{i}\right)U_0(t_{i+1},t_{i})  
\ee  
with $n_r$ kicks for the realization $r$ and  
\be{e38a}  
U_0(t_{i+1},t_{i})=\exp\left(-\frac{i}{\hbar}(t_{i+1}-t_{i})H_0\right)\,.   
\ee  
This representation illustrates how the stochastic driving operates.  Between two kicks at $t_{i}$ and $t_{i+1}$ the molecule evolves freely with $U_0$ according to the Hamiltonian $H_0$ (\eq{e38a}).  At each kicking time $t_i$ the stochastic force induces a phase shift by $\exp\left(\frac{i}{\hbar}x\gamma_{i}\right)$.    
In practice, however, it is easier to compute $|\psi_{r}(t)\rangle$  directly using the Standard FFT Split-operator algorithm~ \cite{feit} with absorbing boundary conditions.   

\begin{figure*} 
\includegraphics[width=12cm]{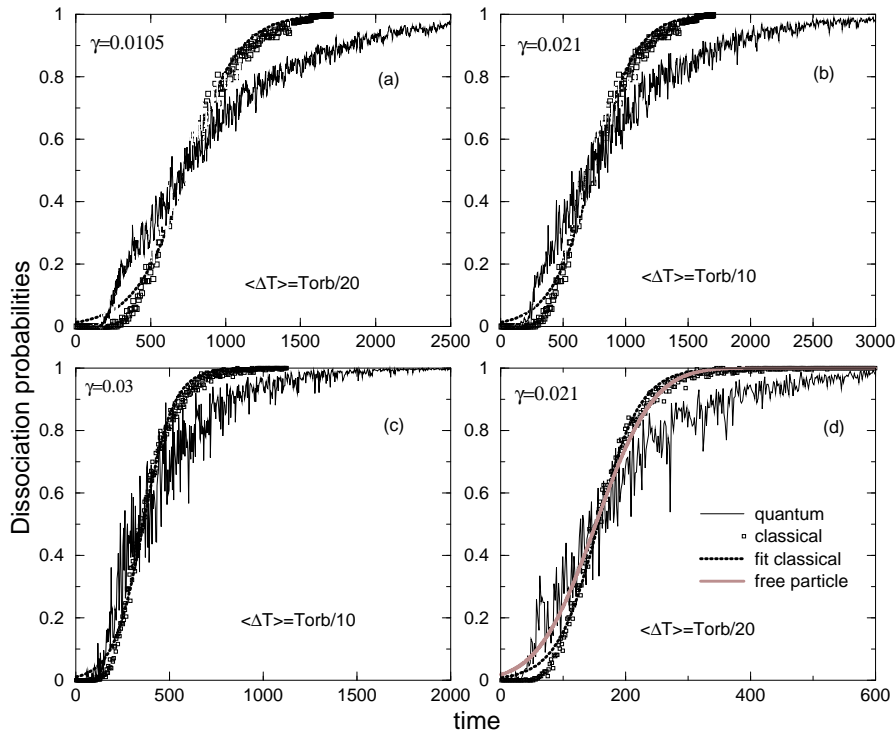}
\caption{Quantum and classical dissociation   probabilities of HF for 
  $\gamma$ and $\langle \Delta t \rangle $ as labelled on each 
  panel, where $T_\mathrm{orb}=0. 085$.  Squares denote the classical
  dissociation probabilities 
 fit with (\ref{fit}).  The red (grey) line is a free-particle model, 
  see text.  Results obtained for HCl are qualitatively similar.}  
\label{Figure2} 
\end{figure*}

\subsection{Dissociation probability}  
The observable we are interested in is the quantum dissociation probability, which is the amount of population in the continuum states.  However, it is easier to calculate the complement, i.~e., the population of all bound states $\psi_\nu$. It reads for a given realization $r$  
\begin{eqnarray} P_{\nu}^{r}(t)=\left|\brkt{\psi_\nu}{\psi_r(t)}\right|^{2}\label{e313}.  \end{eqnarray}  
The dissociation probability for the realization $r$  is then given by  
\begin{eqnarray} P_{d}^r(t)=1- \sum_{\nu=0}^{N_b-1}P_{\nu}^{r}(t)\,. 
\label{e314} \end{eqnarray}  
  
Classically,  $P^r_{d;cl}(t)$ is given in terms of trajectories $N_+(t)$  
which have positive enery $E(t)>0$ at time $t$.   
The physical result is obtained by averaging \eq{e314}  over   
the $N_r$ realizations.  For the results we will present we chose $N_r=100$ and $N_\mathrm{test}=1000$ which was sufficient to achieve convergence.

\section{Results and discussions}  
\subsection{Quantum and classical dissociation probabilities}

An overview over the results is given in Fig.~1.   
As a general trend one sees that quantum and classical dissociation probabilities  
do not coincide, in neither limit of the stochastic force (small and large $\gamma$ and $\langle \Delta t\rangle$).  Furthermore, for all parameter combinations, the classical  
dissociation takes off later but increases faster eventually overtaking the quantum dissociation and going into saturation earlier than in the quantum case.   
The more  abrupt classical dissociation can be parameterized  with 
\begin{eqnarray}  
P_{d;cl}\simeq\frac{1}{2}\tanh(at+b)+\frac{1}{2},\label{fit} \end{eqnarray}  
which fits quite well the classical dissociation.   
The fact, that the discrepancy between the quantum and the classical curves prevails even in the parameter limits for the stochastic force, can be partially attributed to a scaling invariance.   This invariance with respect to the ratio $ \gamma/\langle \Delta t\rangle=\langle F(t)\rangle$ is obeyed by the classical dynamics but not by the  quantum dynamics.  The scaling invariance means that for equal average stochastic force $\langle F(t)\rangle$ (compare \eq{averageforce}),  
the classical dynamics is the same, yet on different effective time  
scales.  This can be seen by transforming the dynamics to a dimensionless time  
variable $\tau = t/ \langle \Delta t\rangle$.   
The effective Hamiltonian in the new time variable $\tau$, $(H-E)dt/d\tau$, remains invariant against changes  
        of $\langle F(t)\rangle=\gamma/\langle \Delta t\rangle$.   
  
While the classical dynamics gives qualitatively the same picture as the quantum dynamics it does not approach the quantum result, not even in the limit of a large number of kicks.  This is different from a Coulomb system under stochastic driving  \cite{burgdorfer}.   
Since it becomes classical close to $E=0$ (which corresponds formally to the dissociation threshold here) one can show that the Coulomb system itself behaves almost classical, and therefore the classical scaling under the stochastic force also applies to the quantum Coulomb system close to $E=0$.  The molecular system behaves non-classically, even close to the dissociation threshold, which prevents to approach the classical scaling under the stochastic force.   
Interestingly, the nature of stochastic driving, namely the cancellation of interferences, does not help to approach classical behavior.  The reason is that the dynamics in the Morse potential without stochastic force differs classically and quantum mechanically, particularly for higher energies, where the non linearity of the potential is strong.  
Consequently, one may ask if under a very strong stochastic force, i. e. , without a potential, classical and quantum dynamics become similar.   
  
\subsection{The free particle limit under stochastic driving} 
If $V=0$ in \eq{hamiltonian}, i. e. , $H_0 = p^2/2m$, one sees immediately with the help of  
Ehrenfest's theorem that classical and quantum observables should agree since \eq{hamiltonian} only contains linear and quadratic operators.  Therefore, the  
 state $|\psi(t)\rangle$ time evolved under the stochastic driving from an initial momentum $|\psi(0)\rangle = |p\rangle$ is simply given by 
 $|\psi(t)\rangle = |p(t)\rangle$ where $p(t)$ is defined by the 
 classical time evolution, starting from the initial momentum $p_i$ at time $t_0$, 
\be{momentum}  
p_t=\lambda\gamma (t-t_0)+ p_i\,.   
\ee 
We can define a formal analog to the dissociation probability, namely 
the probability to find a particle after time $t$ with positive momentum \be{pseudodiss} 
P_d(t) \equiv \int_{0}^\infty d\,p_t W(p_i(p_t))\,, 
\ee 
 
where $p_i(p_t)$ can be obtained from \eq{momentum} and $W(p)$ is the initial momentum distribution which we assume for simplicity to be Gaussian, 
\be{distribution}  
W(p_i)=\alpha/\sqrt{\pi}\exp(-\alpha^2 p_i^2)\,.   
\ee 
Inserting \eq{distribution} and \eq{momentum} into \eq{pseudodiss} leads to an incomplete Gaussian integral with the analytical solution  
\be{errorf}  
P_d(t)=\textstyle\frac{1}{2}\mathrm{Erfc}[-\alpha\lambda\gamma (t-t_0)]\,.   
\ee 
We may use this analytical expression $P_d(t)$ with the two parameters $\alpha, t_0$  to fit and interpret 
the dissociation probabilities in Fig.~1.  
At some time $t_0>0$ after a number of kicks the systems will have a distribution (with width $\alpha$) of energies and the mean energy may be considered to be high enough to approximate the dynamics with a free particle Hamiltonian under stochastic driving without a potential.  
As one can see in Fig.~1, this approximation becomes increasingly better for larger time in comparison to the {\it classical} response, while the quantum response remains different.   
In Fig.~1-d $P_d(t)$ is plotted  for $\gamma=0. 021$, 
$\lambda=\langle \Delta T\rangle^{-1}=235. 301$, $m=1782. 83$, 
$\alpha=0. 02135$ and $t_0=150$.

\section{Conclusions}  
We have proposed and discussed the possibility of  
dissociating diatomic molecules by a stochastic force.  This  
problem has been explored as function of the  
characteristic parameters of the stochastic force, namely the average strength of kicks $\gamma$ and the average time between kicks  $\langle \Delta t\rangle$.   In view of the  
effectivity of the stochastic force to dissociate the molecule  with typical $\langle \Delta t\rangle$ much longer than electronic time scales we expect the stochastic force to be an efficient way to dissociate a molecule.  
In contrast to Coulomb dynamics there is no parameter limit of the stochastic force where classical and quantum results coincide.  The reason is the classical scaling of the dynamics under the stochastic force which is broken by the quantum dynamics.
We recall that the present system is a closed one, not coupled to an environment and therefore not subject to dissipation and diffusion. For the latter case of an open system the classical-quantum correspondence has been investigated systematically,
with the general tendency that strong decoherence makes the quantum system to behave more classically. In contrast, little is known about the quantum-classical correspondence in the present case of a closed system exposed to a stochastic force. 
  
We hope that our results will stimulate efforts to achieve experimental dissociation of diatomic molecules  using white shot noise.  Experiments, using a stochastic force similar to the present one, have been successfully performed by Raizen and cowerkers~\cite{raizen}, on the dynamics of rotors.    
  
\begin{acknowledgments}  
We gratefully acknowledge fruitful discussions with A.  Kolovsky, H.  Kantz, A. R. R.  Carvalho, M. S.  Santhanam, R.  Revaz and A.  Buchleitner.  AK was supported  by the Alexander von Humboldt (AvH) Foundation with the Research fellowship No. IV. 4-KAM 1068533 STP.   
\end{acknowledgments}

%
%
%

\end{document}